\documentclass[aps,pra,onecolumn,groupedaddress,superscriptaddress,tightenlines,showpacs,nofootinbib]{revtex4}
\usepackage{amssymb}
\usepackage{color,graphicx}
\usepackage{lmodern}
\usepackage{amsmath}
\usepackage{amsbsy}
\usepackage{float}
\usepackage{bbm}
\usepackage{bm}
\usepackage{epsfig}
\usepackage{braket}
\usepackage{graphics}
\usepackage{subfigure}
\usepackage{appendix}

\newcommand\D{\operatorname{d}}

\newcommand\tr{\operatorname{tr}}
\newcommand\ad{\operatorname{ad}} 
\newcommand\op{\operatorname{op}}
\newcommand\A{\operatorname{A}}
\newcommand\B{\operatorname{B}}
\newcommand\s{\operatorname{s}}
\newcommand\Int{\operatorname{int}}
\newcommand\I{\operatorname{i}}

\begin{document}
\title{Can Thermodynamic Behavior of Alice's Particle Affect Bob's Particle?}

\author{Ali Soltanmanesh}
\email{soltanmanesh@ch.sharif.edu}
\affiliation{Research Group on Foundations of Quantum Theory and Information,
Department of Chemistry, Sharif University of Technology
P.O.Box 11365-9516, Tehran, Iran}
\affiliation{School of Physics, Institute for Research in Fundamental Sciences (IPM), P.O.Box 19395-5531, Tehran, Iran}

\author{Hamid Reza Naeij}
\email[]{naeij@alum.sharif.edu}
\affiliation{Research Group on Foundations of Quantum Theory and Information,
Department of Chemistry, Sharif University of Technology
P.O.Box 11365-9516, Tehran, Iran}
\author{Afshin Shafiee}
\email[Corresponding Author:~]{shafiee@sharif.edu}
\affiliation{Research Group on Foundations of Quantum Theory and Information,
Department of Chemistry, Sharif University of Technology
P.O.Box 11365-9516, Tehran, Iran}

\begin{abstract}

We propose an experiment to investigate the possibility of long-distance thermodynamic relationships between two entangled particles. We consider a pair of spin-$\frac{1}{2}$ particles prepared in an entangled singlet state in which one particle is sent to Alice and the other to her distant mate Bob, who are spatially separated. Our proposed experiment consists of three different setups: First, both particles are coupled to two heat baths with various temperatures. In the second setup, only Alice's particle is coupled to a heat bath and finally, in the last setup, only Bob's particle is coupled to a heat bath. We study the evolution of an open quantum system using the first law of thermodynamics based on the concepts of ergotropy, adiabatic work, and operational heat, in a quantum fashion. We analyze and compare ergotropy and heat transfer in three setups. Our results show that the heat transfer for each entangled particle is not independent of the thermalization process that occurs for the other one. We prove that the existence of quantum correlations affects the thermodynamic behavior of distant particles in an entangled state.
\end{abstract}
\maketitle
\textbf{keywords}: Quantum thermodynamics, Open quantum system, Long-distance heat transfer, Ergotropy

\section{Introduction}

Quantum thermodynamics is an important growing field of research that focuses on the relations between two physical theories: classical thermodynamics and quantum mechanics. Nowadays, much attention has been dedicated to applications of quantum thermodynamics such as quantum information \cite{Popescu, Sagawa,Del Rio,Brandao} and catalysis \cite{Ng,Lostaglio, Aberg}. Moreover, the study of quantum thermodynamics is currently helping us in development of quantum thermal machines and quantum heat engines, especially in manipulation, management and production of heat and work \cite{Kosloff, Klimovsky,Quan,Roulet}.

The formalism of quantum thermodynamics is connected to the theory of open quantum systems. In other words, in quantum thermodynamics, the heat transfer is explained by a system-bath model. This model is important not only from a fundamental view, but also for the practical applications. Generally, there is a belief that open quantum systems are consistent with laws of equilibrium thermodynamics \cite{Nieuwenhuizen}. In this regard, Binder {\it et al.} formulated an appropriate framework for operational first law of  thermodynamics for an open quantum system undergoing a general quantum process and presented it as a complete positive trace-preserving (CPTP) map \cite{Binder}.

One of the most important challenges in quantum thermodynamics is the investigation of fundamental concepts such as correlation, entanglement and non-locality in related processes. It is believed that the presence of correlation in quantum thermodynamics could be a valuable resource for many quantum information tasks \cite{Sapienza} and a crucial factor in quantum thermal machines and quantum heat engines \cite{Modi}.

In this context, many studies have been done to investigate the foundations of thermodynamics in the quantum domain \cite{Horodecki,Faist,Lostaglio1,Lostaglio2,Dahlsten}. For instance, Chiribella {\it et al.} analyzed the roots of the connection between entanglement and thermodynamics in the framework of general probabilistic theories. They showed that there is a duality between information erasure and entanglement generation \cite{Chiribella}. In quantum thermodynamics, entanglement is also connected to work extraction from multipartite systems \cite{Morris}. Actually, entangling unitary operations are capable of extracting more work than local operations from quantum systems. Alicki and Fannes demonstrated that non-local unitary operations are capable of increasing the amount of work extracted with respect to local operations in a large number of identical copies of a battery \cite{Alicki}. Francica {\it et al.} showed that how the presence of quantum correlations can influence work extraction in the closed quantum systems. They considered a bipartite quantum system and showed that it is possible to optimize the process of work extraction via the concept of ergotropy. They proved that the maximum extracted work is related to the existence of quantum correlations between the two parts of the system \cite{Francica}. Furthermore, heat capacity presented as an indicator of entanglement and investigated the issue of how the entanglement at the ground state of a system affects the third law of thermodynamics \cite{Wiesniak,Rieper}. Other thermodynamic quantities such as magnetic susceptibility \cite{Wiesniak2} and entropy \cite{Bauml} are also proposed as the entanglement witnesses. 

However, the studies in this area have not addressed the possibility of long-distance heat transfer in a quantum fashion. Such a phenomenon can affect the performance of quantum thermal machines, quantum heat engines and related technologies. So, is it possible to observe this behavior in a thermodynamic process?

In the present study, we propose an experiment to investigate the possibility of a bizarre relationship in quantum thermodynamics. For this purpose, we consider a pair of spin-$\frac{1}{2}$ particles prepared in a spin-singlet state. The first particle is sent to Alice and the second one to Bob (the observers), who are far apart. Moreover, we consider three different setups in our proposed experiment: a) both particles in Alice and Bob's sides are coupled to two heat baths with different temperatures, b) only the particle in Alice's side is coupled to a heat bath and c) only the particle in Bob's side is coupled to a heat bath. Then, we study the first law of thermodynamics by using the concepts of ergotropy and adiabatic work in CPTP map in all three scenarios. Our results show that there could be a long-distance thermodynamic relationship between two entangled particles which is responsible for work extraction in a thermodynamic process. 

In the remainder of the paper, we first review what called the operational first law of quantum thermodynamics. Then, we propose an experiment to investigate how the maximum work could be extracted from it. Moreover, we discuss about physical feasibility of our proposed experiment. Finally, the results are discussed in the conclusion section.

\section{Operational First Law of Quantum Thermodynamics}

The first law of thermodynamics explains that the change in the internal energy $(\Delta E)$ for a system consists of two terms: work $(W)$ and heat $(Q)$. Work extraction from a quantum system is a crucial issue in quantum thermodynamics. Therefore, we review some of the concepts necessary to investigate how much work can be extracted from a quantum state in a cyclic unitary evolution.  A cyclic process, here, means that the Hamiltonians of the system at the initial and the end points of the process is identical \cite{Binder}.

We define the internal energy $E$ for the quantum state $\hat{\rho}$ of a system with the Hamiltonian $\hat{H}$ at time $t$ as $E(t)=\tr [\hat{\rho} (t) \hat{H}(t)]$ where $\hat{H}$ is defined as 
\begin{align}
\hat{H}=\sum \varepsilon _n \vert \varepsilon _n \rangle \langle \varepsilon _n\vert \hspace{0.4cm} \text{;} \hspace{0.4cm} \varepsilon _{n+1}\geq \varepsilon _n \forall n
\end{align}

For the quantum state, we can write
\begin{align}
\label{eq2}
\hat{\rho}=\sum r_n \vert r_n\rangle\langle r_n\vert \hspace{0.5cm} \text{;} \hspace{0.4cm} r_{n+1}\leq r_n \forall n
\end{align}

To have the maximum work extraction in a cyclic unitary process, the density matrix of the system in Eq. (2) should end in the states known as passive states $\hat{\pi}$. These states are diagonal in the eigenbasis of $\hat{H}$ with decreasing populations for increasing energy levels, expressed as
\begin{align}
\hat{\pi}=\sum r_n \vert \varepsilon_n\rangle\langle \varepsilon_n\vert \hspace{0.5cm} \text{;} \hspace{0.4cm} r_{n+1}\leq r_n \forall n
\end{align}

We define ergotropy $ \mathcal{W}$, the maximum work that can be extracted from a non-passive state $\hat{\rho}$ concerning  $\hat{H}$ via a cyclic unitary evolution $(\hat{\rho}\longmapsto \hat{\pi})$, as \cite{Binder,Solatnmanesh}
\begin{align}
\mathcal{W}=\tr\big[\hat{\rho}\hat{H}-\hat{\pi}\hat{H}\big]=\sum_{m,n} r_m \varepsilon_n \big[\vert  \langle \varepsilon_n \vert r_m\rangle\vert^2- \delta_{mn}\big]
\end{align}

We now consider a non-cyclic unitary evolution in which the initial and the final Hamiltonians, $\hat{H}$ and $\hat{H}'$, are different where $\hat{H}'=\sum \varepsilon' _n \vert \varepsilon' _n \rangle \langle \varepsilon' _n\vert$ with $\varepsilon' _{n+1}\geq \varepsilon' _n$. We assume that the change in the Hamiltonian from $\hat{H}$ to $\hat{H}'$ is adiabatic. This means that the eigenstates of the Hamiltonian remain unchanged at each instant and the final state $\hat{\pi}'$ will be a passive state with respect to $\hat{H}'$, if the initial state $\hat{\pi}_m$ is passive with respect to $\hat{H}$. This evolution is unitary, so there is no heat transfer and any change in the internal energy $E$ is due to the adiabatic work which can be defined as
\begin{align}
\langle W\rangle_{\ad}=\tr\big[\hat{\pi}'\hat{H}' \big]-\tr \big[\hat{\pi}_m\hat{H}\big]
\end{align}

In a general quantum evolution $(\hat{\rho},\hat{H})\longmapsto (\hat{\rho}',\hat{H}')$, $\Delta E$ is given by
\begin{align}
\label{intE}
\Delta E=\tr\big[\hat{\rho}' \hat{H}'\big]-\tr\big[\hat{\rho} \hat{H}\big]
\end{align}

Considering the concepts of ergotropy and adiabatic work, and defining $\hat{\pi}_m=\sum r'_n \vert \varepsilon_n\rangle\langle\varepsilon_n\vert$, one can show that an operationally meaningful first law of thermodynamics could be introduced as \cite{Binder}
\begin{align}
\label{FL}
\Delta E= \Delta \mathcal{W}+\langle W\rangle_{\ad}+\langle Q\rangle_{\op}
\end{align} 
where $\Delta\mathcal{W}=\mathcal{W}(\hat{\rho}',\hat{H}')-\mathcal{W}(\hat{\rho},\hat{H})$ shows a genuine out of equilibrium contribution and $\langle Q\rangle_{\op}=\tr\big[\hat{\pi}_m \hat{H}\big]-\tr\big[\hat{\pi} \hat{H}\big]$ denotes the heatlike term in $\Delta E$. So, it has been shown that any thermodynamic process that can be portrayed with a CPTP map, obeying Eq. \eqref{FL} introduced as operational first law of quantum thermodynamics (see \cite{Binder}).

\section{The Proposed Experiment}

\begin{figure}
\centering
\begin{subfigure}[]{
\centering
\includegraphics[scale=0.45]{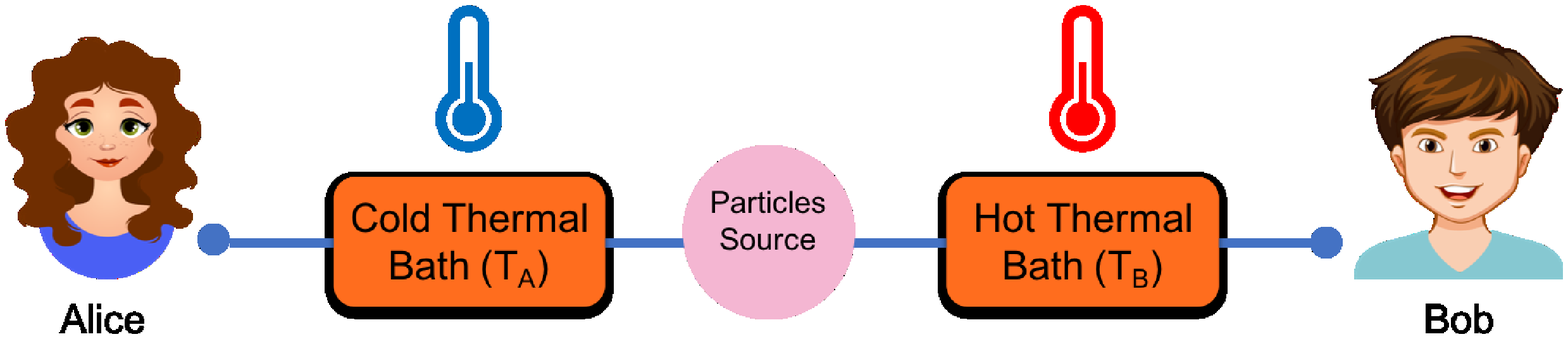}}
\end{subfigure}
\begin{subfigure}[]{
\centering
\includegraphics[scale=0.45]{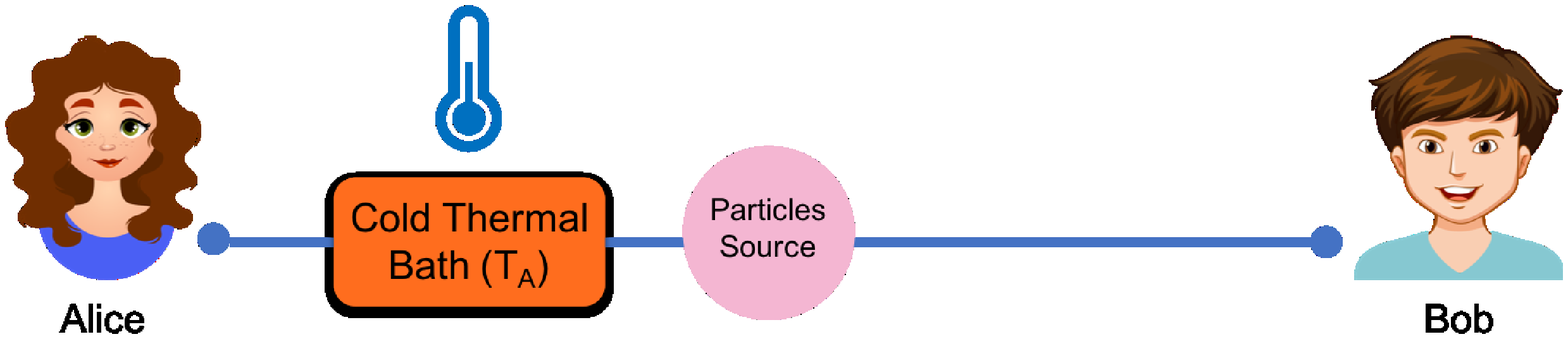}}
\end{subfigure} 
\begin{subfigure}[]{
\centering
\includegraphics[scale=0.45]{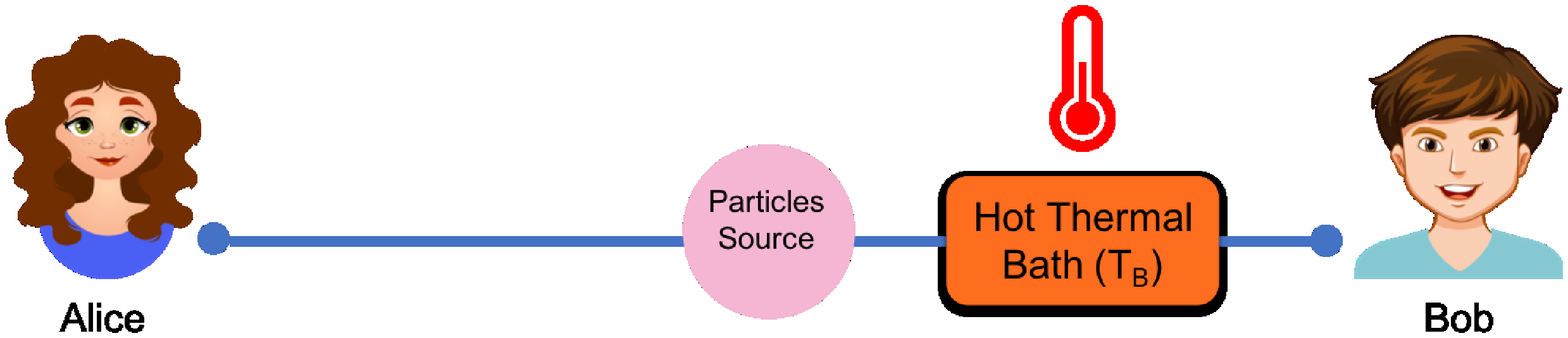}}
\end{subfigure}
\caption{Schematic for our proposed experiment: a) Both particles in Alice and Bob's sides are coupled to two heat baths with  temperatures $T_{\A}$ and $T_{\B}$, respectively, which $T_{\B}> T_{\A}$; b) The particle in Alice's side is coupled to a heat bath with temperature $T_{\A}$; c) The same as (b) but for the observer Bob with thermal bath $T_{\B}$ .}
\end{figure}

Let us consider a pair of spin-$\frac{1}{2}$ particles prepared in a singlet-spin state as
\begin{align}
\label{0state}
\vert\psi\rangle=\frac{1}{\sqrt{2}} \big(\vert +\rangle_{\A}\vert -\rangle_{\B} - \vert -\rangle_{\A}\vert +\rangle_{\B}\big)
\end{align}
where $\hat{\sigma}^{(m)}_z \vert \pm\rangle_m=\pm \vert \pm\rangle_m$, in which $\hat{\sigma}^{(m)}_z$ is the $z$ component of spin Pauli for $\A$ and $\B$ particles where the first particle is sent to Alice and the second one to her distant mate Bob, respectively. Then, we propose three different setups for our proposed experiment as given in FIG. 1.

As is well known, in the system-bath model, the total Hamiltonian can be written as
\begin{align}
\label{Hs}
\hat{H}=\hat{H}_{\s}+\hat{H}_{\varepsilon}+\hat{H}_{\Int}
\end{align}
where $\hat{H}_{\s}$, $\hat{H}_{\varepsilon}$ and $\hat{H}_{\Int}$ are the Hamiltonians of the system, the bath and the system-bath interaction, respectively. We define Hamiltonian of the system as
\begin{align}
\hat{H}_{\s}=\frac{\omega_{\A}}{2} \hat{\sigma}^{(\A)}_z \otimes I+ \frac{\omega_{\B}}{2} I \otimes \hat{\sigma}^{(B)}_z 
\end{align}
where $\omega_{\A}$ and $\omega_{\B}$ are the frequencies of the particles in Alice and Bob's sides, respectively, which we assume that $\omega_{\A} > \omega_{\B}$ and $\hbar=1$. Furthermore, we consider a heat bath consisting of harmonic oscillators as the environment for our three setups. So, we have
\begin{align}
\hat{H}_{\varepsilon}=\sum_{j} \big(\frac{1}{2m_j}\hat{p}^2_j+\frac{1}{2}m_j \Omega^2_j \hat{q}^2_j\big)
\end{align}
where a Bosonic mode $j$ in the bath is described by its frequency $\Omega_j$, mass $m_j$, position $\hat{q}_j$ and momentum $\hat{p}_j$. 

The interaction Hamiltonians for the three setups in FIG. 1 are respectively defined as \cite{Breuer}
\begin{equation}
\hat{H}^{(a)}_{\Int}=-\hat{D}^{(\A)} \otimes \sum_j c_j \hat{q}_j-\hat{D}^{(\B)} \otimes \sum_k c_k \hat{q}_k
\end{equation}
\begin{align}
\hat{H}^{(b)}_{\Int}=-\hat{D}^{(\A)} \otimes \sum_j c_j \hat{q}_j
\end{align}
\begin{align}
\hat{H}^{(c)}_{\Int}=-\hat{D}^{(\B)} \otimes \sum_k c_k \hat{q}_k
\end{align}
In the interaction Hamiltonians, $\hat{D}^{\A(\B)}$ is a dipole operator where can be defined as $\hat{D}^{\A(\B)}={\bf d} \hat{\sigma}^{\A(\B)}_- e^{-i \omega_{{\A}(\B)}t}+{\bf d}^* \hat{\sigma}^{\A(\B)}_+ e^{i \omega_{{\A}(\B)}t}$ in which ${\bf d}$ is the transition matrix element of the dipole operator and $\hat{\sigma}_{\pm}$ are the ladder operators. These Hamiltonians show that each dipole operator for the system is linearly coupled to the position coordinates of harmonic oscillators in the heat bath. Moreover, we consider an initial thermal state for the environment.

We now study the dynamics of the thermalization process in our proposed experiment using a master equation in the Lindblad form in which the norm and positive definiteness of the quantum state are preserved according to CPTP map \cite{Binder, Breuer}. The Lindblad form of master equation for the three setups of the experiment and the solution of them are given in Appendix A. 
 
For the time-independent Hamiltonian of the system, we have $\langle W \rangle_{\ad}=0$ \cite{Binder}. Moreover, our calculations show that $\Delta E=\Delta E_{\A}+\Delta E_{\B}$ where $\Delta E$, $\Delta E_{\A}$ and $\Delta E_{\B}$  are the change in the internal energy for the first, second and third setups, respectively (see Appendix B for the details of calculations of the internal energies). 

Considering Eq. (7), we obtain
\begin{equation}
\label{F}
\Delta \mathcal{W}+\langle Q\rangle_{\op}=\Delta \mathcal{W}_{\A}+\Delta \mathcal{W}_{\B}+\langle Q\rangle_{\op}^{\A}+\langle Q\rangle_{\op}^{\B}
\end{equation}
where $\Delta \mathcal{W} (\langle Q\rangle_{\op})$, $\Delta \mathcal{W}_{\A} (\langle Q\rangle_{\op}^{\A})$ and $\Delta \mathcal{W}_{\B} (\langle Q\rangle_{\op}^{\B})$ are the ergotropy (the operational heat) in the first, second and third setups in FIG. 1, respectively.

After the thermalization process, the particles end up in passive states. Thus, according to the operational first law of thermodynamics Eq. \eqref{FL}, they lose ergotropy and gain operational heat $\langle Q\rangle_{\op}$ during the process. If we consider heat transfer as a {\it local process}, we expect to see that the total heat transfer in the first setup (FIG. 1(a)) should be equal to the sum of heat transfers in the other two setups (FIG. 1(b) and (c)), $\langle Q\rangle_{\op}=\langle Q\rangle_{\op}^{\A}+\langle Q\rangle_{\op}^{\B}$. So, regarding Eq. \eqref{F}, we expect to have
\begin{align}
\label{Main}
\Delta\mathcal{W}(t)=\Delta\mathcal{W}_{\A}(t)+\Delta\mathcal{W}_{\B}(t)
\end{align}
at a definite time $t$. The details of the calculation of $\Delta \mathcal{W}$ for obtaining heat transfer are given in Appendix C. 

Taking into account the definition of ergotropy in Eq. (4), one can show that Eq. (16) can be written as
\begin{align}
\mathcal{W}_{\A}(0)+\mathcal{W}_{\B}(0)-\mathcal{W}(0)+\sum_n \varepsilon_n\big( \langle \varepsilon_n\vert \hat{\rho}-(\hat{\rho}_{\A}+\hat{\rho}_{\B})\vert \varepsilon_n \big) =\sum_n \varepsilon_n\big( \langle \varepsilon_n \vert \hat{\pi}-(\hat{\pi}_{\A}+\hat{\pi}_{\B}) \vert \varepsilon_n \rangle \big)
\end{align}
where $\hat{\rho}(\hat{\pi})$, $\hat{\rho}_{\A}(\hat{\pi}_{\A})$ and $\hat{\rho}_{\B}(\hat{\pi}_{\B})$ denote density matrices (passive states) of the system in the three setups, respectively. Our calculations show that at time near the decoherence time ($\simeq 10^{-7}$s), the left side of Eq. (17) is equal to 
\begin{align}
(\frac{\omega_{\A}-\omega_{\B}}{2})(\frac{\eta _{\B}-1}{2 \bar{n}_{\B}+1}-1)
\end{align}
Also, the right side of Eq. (17) can be obtained as
\begin{align}
(\frac{\omega_{\A}-\omega_{\B}}{2})&\Big[\frac{\eta_{\B}-1}{2 \bar{n}_{\B}+1}+ (2 \bar{n}_{\B}+1) \frac{2\eta_{\B}}{\eta_{\B}-1} +(2 \bar{n}_{\A}+1) \frac{2\eta_{\A}}{\eta_{\A}-1} +(2 \bar{n}_{\A}+1)(2 \bar{n}_{\B}+1) \nonumber\\
&\times \frac{2\eta}{2(\bar{n}_{\B}-\bar{n}_{\A})+(\eta_{\B}-\eta_{\A})+2(\eta_{\B}\bar{n}_{\A}-\eta_{\A}\bar{n}_{\B})}\Big]
\end{align}
where $\eta_{\I}=e^{-\Gamma_{\I}t(2\bar{n}_{\I}+1)}$ ($\I=\A$ or $\B$) and $\eta=\eta_{\A}\eta_{\B}$. Furthermore, $\Gamma_{\I}$ and $\bar{n}_{\I}$ are the decoherence factor and the average number of the particles in the heat bath, respectively (see Appendix A).

\begin{figure}
\centering
\includegraphics[scale=0.6]{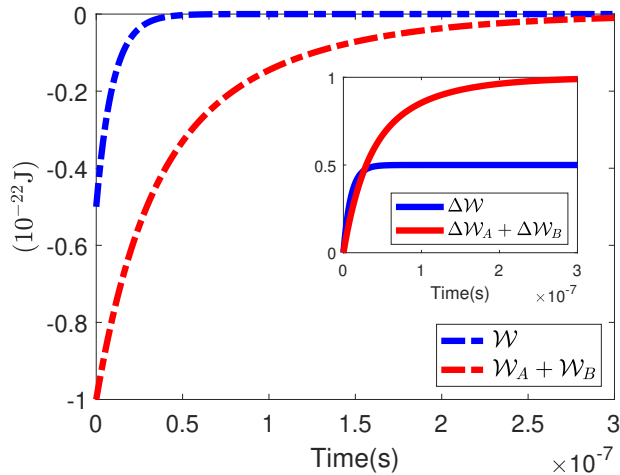}
\caption{The progress of $\mathcal{W}$ and $\mathcal{W}_\text{A}+\mathcal{W}_\text{B}$ (dashed lines) as well as $\Delta\mathcal{W}$ and $\Delta\mathcal{W}_\text{A}+\Delta\mathcal{W}_\text{B}$ (solid lines) with time. As time goes on, the entanglement is eliminated. So, $\mathcal{W}_A+\mathcal{W}_B$ approaches to the value of $\mathcal{W}$. $T_{\A}$ and $T_{\B}$ are considered as $100\text{K}$ and $300\text{K}$, respectively and $\omega_{\A}=2\times 10^{12} {\s}^{-1}$ and $\omega_{\B}=1\times 10^{12} {\s}^{-1}$.} \label{fig2}
\end{figure}

As is clear from Eqs. (18) and (19), Eq. \eqref{Main} does not hold in our proposed experiment (see solid plots in FIG. \ref{fig2}). Therefore, we can simply write $\langle Q\rangle_{\op}\neq\langle Q\rangle^{\A}_{\op}+\langle Q\rangle^{\B}_{\op}$, which means that the heat transfer in the thermalization process of Alice and Bob's baths are not independent from each other. As shown perfectly in dashed plots of FIG. \ref{fig2}, we expect that $\mathcal{W}=\mathcal{W}_{\A}+\mathcal{W}_{\B}$ after decoherence time due to the elimination of entanglement between the particles. Furthermore, from dashed plots in FIG. \ref{fig2}, we conclude that the system loses more ergotropy if the second and third setups are taken together, in comparison to the first setup. Therefore, in application, it is beneficial to extract work in the shape of ergotropy from a single particle rather than entangled ones simultaneously.

It is important to note our results are based on a main relation $\Delta E=\Delta E_{\A}+\Delta E_{\B}$. In other words, we studied our proposed experiment under the operational first law of thermodynamics. We believe that the traditional first law of thermodynamics is incapable of providing a complete description of what really happens in quantum thermodynamic processes. Let us, however, we briefly discuss the situations that are different from our model.
 
1) Pure dephasing model $\Delta E=\Delta E_{\A}=\Delta E_{\B}$=0 \cite{Breuer}: This model is a special case of a general model that we studied. Since system's Hamiltonian is time-independent ($\langle W\rangle_\mathrm{ad}=0$), according to the operational first law of thermodynamics, we have $\Delta\mathcal{W}=-\langle Q\rangle_\mathrm{op}$, $\Delta\mathcal{W}_{\A}=-\langle Q\rangle^{\A}_\mathrm{op}$ and $\Delta\mathcal{W}_{\B}=-\langle Q\rangle^{\B}_\mathrm{op}$. Therefore, if we consider the heat flow as a local process, we must have $\langle Q\rangle_\mathrm{op}=\langle Q\rangle^{\A}_\mathrm{op}+\langle Q\rangle^{\B}_\mathrm{op}$ and subsequently $\Delta\mathcal{W}=\Delta\mathcal{W}_{\A}+\Delta\mathcal{W}_{\B}$.  When $\Delta\mathcal{W}\neq\Delta\mathcal{W}_{\A}+\Delta\mathcal{W}_{\B}$, we expect that the heat flow of the entangled particles are dependent to each other. So, the traditional first law of thermodynamics cannot describe such a system.
 
 2) Other types of system-bath interactions with $\Delta E=\Delta E_{\A}+\Delta E_{\B}$:  According to the operational first law of thermodynamics, for any system bath interaction that reaches this same expression for the internal energy, if decrease in ergotropy is different for different system-bath interaction, we have $\langle Q\rangle_\mathrm{op}\neq\langle Q\rangle^{\A}_\mathrm{op}+\langle Q\rangle^{\B}_\mathrm{op}$. So, we expect to observe the heat flow of the entangled particles is dependent to each other.
 
 3) In a case with $\Delta E\neq\Delta E_{\A}+\Delta E_{\B}$: Such case had not been observed or reported. Also, we cannot name a system with this behavior. However, if such a case exists, it would be quite interesting. In this case for entangled particles with different changes in ergotropy for different system-bath interactions, we cannot predict how heat flow changes. Nevertheless, any energy transfer being dependent on different system-bath interactions for spatially entangled particles would be even more bizarre. If such cases exist, it will directly show long-distance energy transfers, independent of the first law of thermodynamics. 
 
 In the case of pure dephasing models in which internal energy does not change, the traditional first law of thermodynamics provides no description of what happens in the process. However, the operational form of the first law completely describes such systems using the concept of ergotropy \cite{Binder}. Interestingly, the pure dephasing model is a special case of the general model that we discuss. Moreover, so many studies showed that the traditional well-known heat is inconsistent with the thermodynamics of quantum systems, Clausius inequality and quantum second law of thermodynamics based on information theory \cite{Nieuwenhuizen,Solatnmanesh,Allahverdyan,Hilt}.

We should note that despite that we can gain work at the cost of losing coherences, however in this work, coherence alone is not enough, and the presence of entanglement is essential for reaching the main result of this work for spatially separated particles. For the first setup, in a case that entanglement is not included, for the two particles A and B, that has been sent to Alice and Bob sides, respectively, we always can separate the initial density matrix as $\hat{\rho}(0)=\hat{\rho}_{\A}(0)\otimes \hat{\rho}_{\B}(0)$. Since there are no correlations between the particles (they are spatially separated and there is no interaction between particles in the Hamiltonian), they remain separated and not entangled in all times. Thus, we have $ \hat{\rho}(t)=\hat{\rho}_{\A}(t)\otimes \hat{\rho}_{\B}(t)$. Now, regarding $\hat{H}_s=\hat{H}_{\A}+\hat{H}_{\B}$ we can calculate the ergotropy in all times as
\begin{align}
\mathcal{W}&=\tr [\hat{\rho} \hat{H}_s-\hat{\pi} \hat{H}_s]=\tr [\hat{\rho}_{\A}\hat{H}_{\A}\otimes \hat{\rho}_{\B}+\hat{\rho}_{\A}\otimes \hat{\rho}_{\B}\hat{H}_{\B} -\hat{\pi}_{\A}\hat{H}_{\A}\otimes \hat{\pi}_{\B}-\hat{\pi}_{\A}\otimes \hat{\pi}_{\B}\hat{H}_{\B}] \nonumber \\
&=\tr[\hat{\rho}_{\A}\hat{H}_{\A}-\hat{\pi}_{\A}\hat{H}_{\A}]+\tr[\hat{\rho}_{\B}\hat{H}_{\B}-\hat{\pi}_{\B}\hat{H}_{\B}]=\mathcal{W}_{\A}+\mathcal{W}_{\B}
\end{align}

As we can see in a case without an entanglement, we always have $\Delta\mathcal{W}=\Delta\mathcal{W}_{\A}+\Delta\mathcal{W}_{\B}$ that is in a complete agreement with Eq. \eqref{Main} and  system's behavior differs with the case that entanglement is included and $\Delta\mathcal{W}\neq\Delta\mathcal{W}_{\A}+\Delta\mathcal{W}_{\B}$

\begin{figure}
\centering
\includegraphics[scale=0.45]{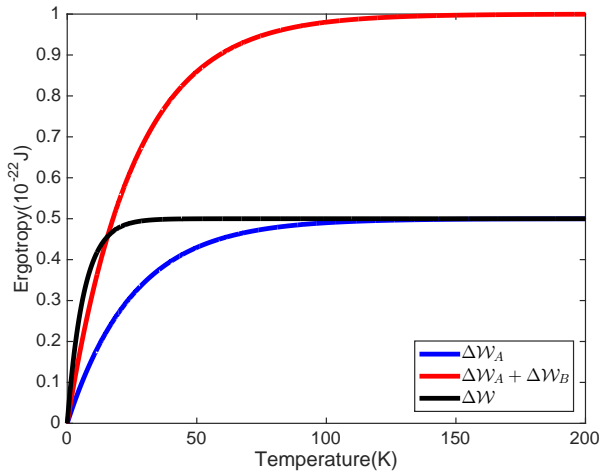}
\caption{The progress of $\Delta\mathcal{W}$, $\Delta\mathcal{W}_{\A}$ and $\Delta\mathcal{W}_{\A}+\Delta\mathcal{W}_{\B}$ with temperatures of Alice's and Bob's baths, in which the temperatures of the baths are the same.}\label{fig3}
\end{figure}

\begin{figure}
\centering
\includegraphics[scale=0.45]{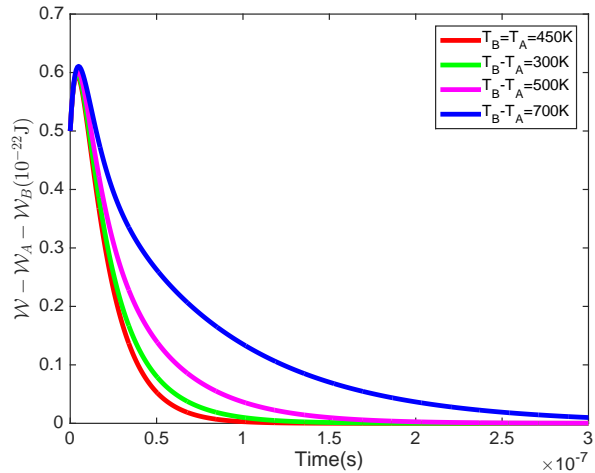}
\caption{The progress of $\mathcal{W}-\mathcal{W}_\text{A}-\mathcal{W}_\text{B}$ with time in a variety of temperature differences between two heat baths, $T_{\B}-T_{\A}$. For all plots $(T_{\A}+T_{\B})/2=450\text{K}$ and $\omega_{\A}=2\times 10^{12} {\s}^{-1}$ and $\omega_{\B}=1\times 10^{12} {\s}^{-1}$.}\label{fig4}
\end{figure}

For a better realization of the system's behavior, we plotted the changes of ergotropy versus the temperatures of the baths for the first setup, in which the temperatures of Alice's and Bob's baths are the same. Interestingly, FIG. \ref{fig3} clearly shows that the entangled system interacting with two baths with the same temperature tends to behave similarly to the system interacting with a single bath. Also, FIG. \ref{fig3} shows that with a higher temperature of thermal baths, the system loses more ergotropy during the thermalization process. In FIG. \ref{fig2} and FIG. \ref{fig3}, we observe that the progress of $\Delta\mathcal{W}$ is almost equivalent to $\Delta\mathcal{W}_{\A}+\Delta\mathcal{W}_{\B}$ {\it until} the state of the system thermalizes and the changes in ergotropy tend to the single bath setup. Accordingly, the interaction with the bath with a higher temperature is the limiting factor for the thermalization of the state of the system and affect the other particle-bath interaction. This limiting factor causes that in a setup with more differences in temperatures of the baths, we see more violations from Eq. \eqref{Main} rather than the case with baths in the same temperatures, as is shown in FIG. \ref{fig4}.

\section{Physical Feasibility of Our Proposed Experiment}

One of the most important challenges for the realization of our proposed experiment is the preparation of the singlet-spin state of a pair of spin-$\frac{1}{2}$ particles. In this regard, for example, the gate voltage control of the exchange interaction is used to prepare, manipulate, and measure two-electron spin states in a gallium arsenide (GaAs) double quantum dot. By placing two electrons in a single dot at low temperatures, the system is prepared in a singlet-spin state. The singlet-spin state is spatially separated by transferring an electron to an adjacent dot \cite{Petta}. Moreover, rapid adiabatic passage (RAP) can prepare an entangled singlet-spin state of two electrons \cite{Taylor}. RAP is a process that permits the transfer of a population between two applicable quantum states. This broad-spread technique originated in nuclear magnetic resonance but is used in virtually all fields such as laser-chemistry, modern quantum optics and quantum information processing \cite{Chen}.
  
Another critical challenge is the heat bath preparation. In this regard, a thermal source prepared by collecting emitted light of ionization processes in hot vapor of atoms like mercury and hydrogen \cite{Ou}. Passing laser beams through a fast rotating grounded glass is another useful way of making thermal sources \cite{Rousseau, Estes, Tirandaz}. Moreover, the electromagnetic field in an optical cavity can be considered as a bath. The field in the cavity can be prepared in the state we are interested in using the combination of the various frequencies of the field. It has been shown that the heat bath at different temperatures can be simulated by a careful superposition of bremsstrahlung spectra with different endpoint energies \cite{Mohr}. Harrington {\it et al.} showed that the heat bath can be created by mixing low-frequency noise up to near the qubit frequency with single-sideband modulation \cite{Harrington}. The laser cooling techniques are also used to generate a heat bath \cite{Marzoli,Poyatos}. Furthermore, laser detuning techniques and nanomechanical heat engines are contemporary methods for regulating the temperature of heat baths in desired limits \cite{Dechant, Blickle}.

To establish our thought experiment in the laboratory, after the preparation of the initial state, one can use the method introduced by Skrzypczyk {\it et al.} to simulate and study the system-bath interaction \cite{Skrzypczyk}. In this protocol, we are going through two stages: First, we transform the state of the system into a passive state without using the thermal bath. In the second stage, we transform the state of the system to a thermal state using the qubits of a thermal bath one by one.

In the first stage, we have our system $\hat{\rho}_s$ and a work storage device (a weight that can be raised or lowered) $\hat{\rho}_w$ initially in an uncorrelated state with the density operator $\hat{\rho}_s\otimes \hat{\rho}_w$. We expand $\hat{\rho}_s$ in terms of its eigenvalues $r_n$ and eigenvectors $\vert r_n\rangle$ in which $r_{n+1}\leq r_n$, according to Eq. \eqref{eq2}. Keeping in mind the energy eigenstates $\vert \varepsilon_n\rangle$ with corresponding eigenvalues $\varepsilon_n$, we apply the unitary transform
\begin{align}
\hat{V}=\sum_n\vert \varepsilon_n\rangle\langle r_n\vert\otimes \hat{\Gamma}_{\epsilon_n}
\end{align}
where $\epsilon_n=\langle r_n\vert \hat{H}_s\vert r_n\rangle-\varepsilon_n$ and $\hat{\Gamma}_{\epsilon_n}$ is the translation operator which acts on the position states of the weight as $\hat{\Gamma}_{\epsilon_n}\vert x\rangle=\vert x+\epsilon_n\rangle$. In this manner, $\hat{V}$ always conserves average energy. After applying the transformation, we have the final state as
\begin{align}
\hat{\sigma}_{sw}=\sum_n r_n \vert \varepsilon_n\rangle\langle\varepsilon_n\vert\otimes\hat{\Gamma}_{\epsilon_n}\hat{\rho}_w\hat{\Gamma}^\dagger_{\epsilon_n}
\end{align}
where the reduced state of the system is $\hat{\sigma}_s=\tr_w[\hat{\sigma}_{sw}]$, which is a passive state $\hat{\sigma}_s=\hat{\pi}_s$, and the reduced state of the weight is $\hat{\sigma}_w=\tr_s[\hat{\sigma}_{sw}]$. The change in the average energy of the system is $\Delta E=\tr[\hat{\pi}_s \hat{H}_s]-\tr[\hat{\rho}_s \hat{H}_s]$ which is equal to the change of the ergotropy of the system (Eq. (4)). Moreover, when the entropy of the system remains unchanged, the work extracted as ergotropy is equal to the free energy lost ($\Delta F$) by the system. So, we have
\begin{align}
\Delta F=F(\hat{\rho}_s)-F(\hat{\sigma}_s)=\Delta\mathcal{W}_s
\end{align}

In this stage, an experimentalist measures the free energy of the system, or the amount of work that stored in the weight to see the changes in ergotropy of the system for each setup.

In the second stage, in order to transform the passive state to a thermal state, one can apply $N$ steps, in each step transforming the state $\hat{\pi}_s=\sum_n r_n\vert \varepsilon_n\rangle\langle\varepsilon_n\vert$ into a new passive state $\hat{\pi}'_s=\sum_n r'_n\vert \varepsilon_n\rangle\langle\varepsilon_n\vert$, in which $r'_1=r_1+\delta r$ and $r'_0=r_0-\delta r$ in a situation that $\vert\delta r\vert\ll r_1\leq r_0$. To achieve the mentioned transformation, we take a qubit from a thermal bath with energy eigenstates as

\begin{align}
\hat{\rho}_\mathrm{Bath}=\frac{r'_0}{r'_0+r'_1}\vert 0\rangle\langle 0\vert+\frac{r'_1}{r'_0+r'_1}\vert 1\rangle\langle 1\vert,
\end{align}
where the ratio of the ground and excited state population matches with the states in $\hat{\pi}'_s$. Then, we apply the unitary transformation that swaps this qubit with the state of the system and translate the energy to weight for energy conservation. In our case of two-dimensional sub-space, this transformation maps
\begin{align}
\vert\varepsilon_0\rangle_s\vert 1\rangle_\mathrm{Bath}\vert x\rangle_w \rightarrow\vert \varepsilon_1\rangle_s\vert 0\rangle_\mathrm{Bath}\vert x+\epsilon\rangle_w
\end{align}
 where for all $x$, $\epsilon=E_\mathrm{Bath}-(\varepsilon_1-\varepsilon_0)$. This transformation leaves the system in the state $\hat{\pi}'_s$. We should note that this unitary transformation commutes with the total Hamiltonian. Performing a sequence of $N$ same steps, interacting a system with a new thermal qubit in each step, makes the system ends up in a desired thermal state in $T$.

 \section{Conclusion}

The effects of quantum features of a system on its thermodynamic performance appear currently to be a quite controversial issue. In this regard, many studies have been done to investigate the role of quantum correlations on applications of quantum thermodynamics such as quantum information, quantum thermal machines and quantum heat engines. Most works are now focusing on quantum correlations to improve the efficiency of quantum thermal machines. However, they have not addressed the possibility of the long-distance energy exchanges in the thermodynamic processes.

In this work, we proposed an experiment to investigate the possibility of a distant thermodynamic relationship between a pair of spin-$\frac{1}{2}$ particles prepared in a spin-singlet state in a system-bath interaction model. Our results show that the heat transfer for each entangled particle is not independent of the thermalization process that occurs for the other one. We proved that the existence of quantum correlations affects the thermodynamic behavior of entangled particles that are spatially separated. We showed that the system loses more ergotropy if the second and third setups are taken together, in comparison to the first setup in our proposed experiment. Furthermore, we fully explained the physical feasibility of the experiment in the laboratory.

There is a general belief that entanglement in quantum thermodynamics is related to work extraction from multipartite systems. In other words, entangling operations can extract more work than local operations from quantum systems. Also, the power and entanglement are related to each other in the work extraction process. The correlations could store and extract work. Moreover, it has been proved that the correlations in the forms of quantum mutual information and entanglement can affect the operation of thermal machines and heat engines.

According to this, the observation of distant thermodynamic relationships due to quantum correlations in this study affects the work extraction, the power generation and the cyclical or non-cyclical operations of thermal machines, such as engines and refrigerators. Especially, in this way, it could be possible to extract more work from thermal machines to enhance their efficiency and performance. Moreover, the correlations in quantum thermodynamics are very important in quantum information researches such as investigation of quantum discord and clarifying the concept of entropy and also relevant in thermodynamic transformations. Another important topic in quantum technology is charging a quantum battery including single and two-qubit battery. It is shown that the quantum correlations affect the process of charging a battery, i.e., storing energy in a quantum system for later use. Since all spin-spin $\frac{1}{2}$ interactions are possible to implement with trapped ions, the theoretical predictions for extracting ergotropy for the qubit battery is testable with current experimental techniques. Furthermore, entanglement between the subsystems is understood to play an important role in the kinematic description of thermalization processes.

The present study describes the fundamental concept of a thermodynamic holistic behavior for entangled quantum systems and explains the concept of discrepancy in thermodynamics of parts of an entangled quantum system in a new approach. Moreover, extracting work in the shape of ergotropy from an entangled quantum system may affect on new technological potentials of quantum thermodynamics.

We hope that experimental research groups will evaluate the results of our thought experiment in the future.

\appendix
\section{The Lindblad master equation}

We study the evolution of an open quantum system by using the master equation in the Lindblad form \cite{Breuer}. We first consider an Ohmic spectral density function for the frequencies of the particles of the heat bath with Lorentz-Drude form defined as \cite{Schlosshauer}
\begin{align}
J(\Omega)=\frac{2 \gamma_0 \Omega}{\pi}\frac{\Lambda^2}{\Lambda^2+\Omega^2}
\end{align}
where $\Lambda$ is the cut-off frequency and $\gamma_0$ is a dimensionless factor which describes the system-bath coupling strength.

For the first setup described in FIG. 1, the master equation in the Lindblad form can be written as
\begin{widetext}
\begin{align}
\frac{\D}{\D t}\hat{\rho}(t)&=\Gamma _{\A}(\bar{n}_{\A}+1)\big[\hat{\sigma}_-\otimes I \hat{\rho} \hat{\sigma}_-\otimes I-\frac{1}{2}\hat{\sigma}_+\hat{\sigma}_-\otimes I \hat{\rho}-\frac{1}{2}\hat{\rho}\hat{\sigma}_+\hat{\sigma}_-\otimes I\big] \nonumber\\
&+\Gamma _{\B}(\bar{n}_{\B}+1)\big[I \otimes \hat{\sigma}_-\hat{\rho} I\otimes \hat{\sigma}_+ -\frac{1}{2}I \otimes \hat{\sigma}_+\hat{\sigma}_-\hat{\rho}-\frac{1}{2}\hat{\rho} I \otimes \hat{\sigma}_+\hat{\sigma}_-\big] \nonumber\\
&+\Gamma _{\A}\bar{n}_{\A}\big[ \hat{\sigma}_+ \otimes I \hat{\rho} \hat{\sigma}_- \otimes I-\frac{1}{2} \hat{\sigma}_-\hat{\sigma}_+\otimes I \hat{\rho}-\frac{1}{2} \hat{\rho}\hat{\sigma}_-\hat{\sigma}_+\otimes I\big] \nonumber\\
&+\Gamma _{\B}\bar{n}_{\B}\big[ I \otimes \hat{\sigma}_+ \hat{\rho} I \otimes \hat{\sigma}_- -\frac{1}{2} I \otimes \hat{\sigma}_-\hat{\sigma}_+ \hat{\rho} -\frac{1}{2} \hat{\rho}I\otimes \hat{\sigma}_-\hat{\sigma}_+ \big] 
\end{align}
\end{widetext}
where $\Gamma_{\I}=\gamma^2_0 \omega_{\I} r^2_{\I} \vert {\bf d}\vert ^2/ (1+r^2_{\I})$ and $\bar{n}_{\I}=(e^{\omega_{\I}/K T_{\I}}-1)^{-1}$ are the decoherence factor and the average number of the environmental particles, respectively and $r_{\I}=\Lambda_{\I}/\omega_{\I}$ (i=A or B). By solving the master equation in Eq. (A2), the state of the system can be obtained as
\begin{align}
\label{rho1}
\hat{\rho}(t)=
\begin{pmatrix}
  \lambda_{11} & 0 & 0 & 0 \\
0 & \lambda_{22} & \lambda_{23} & 0\\
0 & \lambda_{32} & \lambda_{33} & 0\\
0 & 0 & 0 & \lambda_{44}
 \end{pmatrix}
\end{align}
where we define
\begin{widetext}
\begin{align}
& \lambda_{11} =\frac{1}{2(2\bar{n}_{\A}+1)(2\bar{n}_{\B}+1)}\big[ \eta_{\B}\bar{n}_{\A}-\eta \bar{n}_{\A}+\eta_{\A} \bar{n}_{\B}-\eta \bar{n}_{\B}-2\eta \bar{n}_{\A}\bar{n}_{\B}+2\bar{n}_{\A}\bar{n}_{\B}\big] \nonumber\\
&\lambda_{22}= \frac{1}{2(2\bar{n}_{\A}+1)(2\bar{n}_{\B}+1)}\big[ \eta_{\A}-\eta_{\B}\bar{n}_{\A}+\eta \bar{n}_{\A}+\eta_{\A}\bar{n}_{\B}+\eta \bar{n}_{\B}+2\eta \bar{n}_{\A}\bar{n}_{\B}+2\bar{n}_{\A}+2\bar{n}_{\A}\bar{n}_{\B}\big] \nonumber\\
&\lambda_{23}=\lambda_{32}=\frac{-\eta}{2} \nonumber\\
&\lambda_{33}=\frac{1}{2(2\bar{n}_{\A}+1)(2\bar{n}_{\B}+1)}\big[\eta_{\B}-\eta_{\A}\bar{n}_{\B}+\eta \bar{n}_{\B}+\eta_{\B}\bar{n}_{\A}+\eta \bar{n}_{\A}+2\eta \bar{n}_{\A}\bar{n}_{\B}+2\bar{n}_{\B}+2\bar{n}_{\A}\bar{n}_{\B}\big] \nonumber\\
&\lambda_{44}=\frac{1}{2(2\bar{n}_{\A}+1)(2\bar{n}_{\B}+1)}\big[-\eta_{\A}-\eta_{\B}-\eta_{\B}\bar{n}_{\A}-\eta \bar{n}_{\A}-\eta_{\A}\bar{n}_{\B}-\eta \bar{n}_{\B}-2\eta \bar{n}_{\A}\bar{n}_{\B}+2\bar{n}_{\A}+2\bar{n}_{\B}+2\bar{n}_{\A}\bar{n}_{\B}+2\big]
\end{align}
\end{widetext}
where $\eta_{\A}=e^{-\Gamma_{\A}t(2\bar{n}_{\A}+1)}$, $\eta_{\B}=e^{-\Gamma_{\B}t(2\bar{n}_{\B}+1)}$ and $\eta=\eta_{\A}\eta_{\B}$.

For the second setup, the master equation can be written as
\begin{widetext} 
\begin{align}
\frac{\D}{\D t}\hat{\rho}_{\A}(t)&=\Gamma_{\A}(\bar{n}_{\A}+1)\big[\hat{\sigma}_- \otimes I\hat{\rho}_{\A} \hat{\sigma}_+\otimes I-\frac{1}{2} \hat{\sigma}_+\hat{\sigma}_-\otimes I \hat{\rho}_{\A}-\frac{1}{2} \hat{\rho}_{\A} \hat{\sigma}_+\hat{\sigma}_-\otimes I\big] \nonumber\\
&+\Gamma_{\A}\bar{n}_{\A}\big[\hat{\sigma}_+\otimes I \hat{\rho}_{\A} \hat{\sigma}_-\otimes I-\frac{1}{2} \hat{\sigma}_-\hat{\sigma}_+ \otimes I \hat{\rho}_{\A}-\frac{1}{2} \hat{\rho}_{\A} \hat{\sigma}_-\hat{\sigma}_+\otimes I \big]
\end{align}
\end{widetext}

By solving the master equation, we have
\begin{align}
\label{rhoA}
\hat{\rho}_{\A}(t)=
\begin{pmatrix}
  \lambda^{\A}_{11} & 0 & 0 & 0 \\
0 & \lambda^{\A}_{22} & \lambda^{\A}_{23} & 0\\
0 & \lambda^{\A}_{32} & \lambda^{\A}_{33} & 0\\
0 & 0 & 0 & \lambda^{A}_{44}
 \end{pmatrix}
\end{align}
where
\begin{align}
&\lambda^{\A}_{11}=\frac{\bar{n}_{\A}-\eta_{\A}\bar{n}_{\A}}{2(2\bar{n}_{\A}+1)}~~~\lambda^{\A}_{22}=\frac{\eta_{\A}+\bar{n}_{\A}+\eta_{\A}\bar{n}_{\A}}{2(2\bar{n}_{\A}+1)} \nonumber\\
& \hspace{0.95in} \lambda^{\A}_{23}=\lambda^{\A}_{32}=\frac{-\eta^{1/2}_{\A}}{2}\nonumber\\
&\lambda^{\A}_{33}=\frac{\bar{n}_{\A}+\eta_{\A}\bar{n}_{\A}+1}{2(2\bar{n}_{\A}+1)}~~~\lambda^{\A}_{44}=\frac{\bar{n}_{\A}-\eta_{\A}-\eta_{\A}\bar{n}_{\A}+1}{2(2\bar{n}_{\A}+1)}
\end{align}

For the third setup, the results of the master equation are similar to the second setup.

\section{Calculation of the changes in internal energies}
\setcounter{equation}{0}

According to Eq. \eqref{intE}, we can calculate the changes in internal energies for the three setups as follows, respectively:
\begin{align}
\label{Eab}
\Delta E(t)&=\mathrm{tr}[\hat\rho(t)\hat{H_s}]-\mathrm{tr}[\hat\rho(0)\hat{H_s}] \\
\label{Ea}
\Delta E_\mathrm{A}(t)&=\mathrm{tr}[\hat\rho_\mathrm{A}(t)\hat{H_s}]-\mathrm{tr}[\hat\rho_\mathrm{A}(0)\hat{H_s}] \\
\label{Eb}
\Delta E_\mathrm{B}(t)&=\mathrm{tr}[\hat\rho_\mathrm{B}(t)\hat{H_s}]-\mathrm{tr}[\hat\rho_\mathrm{B}(0)\hat{H_s}].
\end{align}

One can simply show that the initial internal energies are $\mathrm{tr}[\hat\rho(0)\hat{H_s}]=\mathrm{tr}[\hat\rho_\mathrm{A}(0)\hat{H_s}]=\mathrm{tr}[\hat\rho_\mathrm{B}(0)\hat{H_s}]=0$, Considering Eq. \eqref{0state} as the initial state, Eq. \eqref{Hs} as the Hamiltonian of the system and Eq. \eqref{rho1} as the final state, we obtain $\Delta E$ for the first setup as

\begin{align}
\label{DE}
\Delta E=\lambda_{11}(\frac{\omega_{\A}+\omega_{\B}}{2})+\lambda_{22}(\frac{\omega_{\A}-\omega_{\B}}{2}) +\lambda_{33}(\frac{\omega_{\B}-\omega_{\A}}{2})+\lambda_{44}(\frac{-\omega_{\A}-\omega_{\B}}{2}) =\frac{\omega_{\A}(\eta_{\A}-1)}{2(2\bar{n}_{\A}+1)}+\frac{\omega_{\B}(\eta_{\B}-1)}{2(2\bar{n}_{\B}+1)}.
\end{align}

In the second setup, by using Eq. \eqref{rhoA} as the final state, we have

\begin{align}
\label{DEA}
\Delta E_{\A}=\lambda_{11}^{\A}(\frac{\omega_{\A}+\omega_{\B}}{2})+\lambda_{22}^{\A}(\frac{\omega_{\A}-\omega_{\B}}{2}) +\lambda_{33}^{\A}(\frac{\omega_{\B}-\omega_{\A}}{2})+\lambda_{44}^{\A}(\frac{-\omega_{\A}-\omega_{\B}}{2}) =\frac{\omega_{\A}(\eta_{\A}-1)}{2(2\bar{n}_{\A}+1)}.
\end{align}
Finally, for the third setup with the similar calculations to the second one, we have
\begin{align}
\label{DEB}
\Delta E_{\B}=\frac{\omega_{\B}(\eta_{\A}-1)}{2(2\bar{n}_{\B}+1)}.
\end{align}
Now, one can show that the changes in internal energy for the first setup is equal to the sum of the internal energy changes in the other two:
\begin{align}
\Delta E=\Delta E_{\A}+\Delta E_{\B}
\end{align}

\section{Calculation of ergotropy}
\setcounter{equation}{0}
 
Considering Eq. (15), we calculate ergotropy that can be extracted from the three setups of our experiment, when $T_{\B} > T_{\A}$ and $\omega_{\A} > \omega_{\B}$. 

For the first setup, one obtains 
\begin{align}
\mathcal{W}(t)=(\lambda_{22}+\lambda_{33}+\Delta)(\frac{\omega_{\A}-\omega_{\B}}{2})(\frac{-1}{1+\beta^2})+(\lambda_{22}+\lambda_{33}-\Delta)(\frac{\omega_{\A}-\omega_{\B}}{2})(\frac{\alpha^2}{1+\alpha^2})
\end{align}
where 
\begin{align}
&\alpha=\frac{\lambda_{22}-\lambda_{33}-\Delta}{2\lambda_{32}} \nonumber\\
&\beta=\frac{\lambda_{22}-\lambda_{33}+\Delta}{2\lambda_{32}} \nonumber\\
&\Delta=\big((\lambda^2_{33}-\lambda^2_{22})^2+4\lambda_{23}\lambda_{32}\big)^{1/2}
\end{align}

For the second setup, ergotropy can be obtained as
\begin{align}
\mathcal{W}_{\A}(t)=(\lambda^{\A}_{22}+\lambda^{\A}_{33}-\Delta^{\A})(\frac{\omega_{\A}-\omega_{\B}}{2})(\frac{\alpha^2_{\A}}{1+\alpha^2_{\A}}) +(\lambda^{\A}_{22}+\lambda^{\A}_{33}+\Delta^{\A})(\frac{\omega_{\A}-\omega_{\B}}{2})(\frac{-1}{1+\beta^2_{\A}})
\end{align}
where
\begin{align}
&\alpha_{\A}=\frac{\lambda^{\A}_{22}-\lambda^{\A}_{33}-\Delta^{\A}}{2\lambda^{\A}_{32}} \nonumber\\
&\beta_{\A}=\frac{\lambda^{\A}_{22}-\lambda^{\A}_{33}+\Delta^{\A}}{2\lambda^{\A}_{32}} \nonumber\\
&\Delta^{\A}=\big((\lambda^{\A}_{33}-\lambda^{\A}_{22})^2+4\lambda^{\A}_{23}\lambda^{\A}_{32}\big)^{1/2}
\end{align}

For the third setup, the results are similar to the second setup.

\textbf{Author Contributions} 

A.So proposed the idea and did the calculations, H.R.N and A.Sh contributed to the development and completion of the idea, analyzing the results and discussions. A.So, H.R.N and A.Sh participated in writing the manuscript.

\textbf{Competing Financial Interests}

The authors declare no competing financial interests.
\end{document}